\documentclass[12pt]{article}
\usepackage{amsmath}
\usepackage{feynmf}
\usepackage{latexsym}
\usepackage{amssymb}
\usepackage{graphicx}
\usepackage{caption}
\usepackage{rotating}
\usepackage{latexsym}
\usepackage{verbatim}
\usepackage{epsfig}
\usepackage{multirow}
\def\be{\begin{equation}}
\def\ee{\end{equation}}
\newcommand{\benn}{\begin{equation*}}
\newcommand{\eenn}{\end{equation*}}
\newcommand{\beann}{\begin{eqnarray*}}
\newcommand{\eeann}{\end{eqnarray*}}
\def\bea{\begin{eqnarray}}
\def\eea{\end{eqnarray}}
\def\bei{\begin{itemize}}
\def\eei{\end{itemize}}
\def\bs{\begin{slide}}
\def\es{\end{slide}}
\def\nn{\nonumber}

\def\L{\mathcal{L}}

\def\a{\alpha}

\def\({\left(}
\def\){\right)}
\def\[{\left[}
\def\]{\right]}

\def\Tr{\textnormal{Tr}}

\newcommand{\bth}{{\bf 3}}
\newcommand{\btw}{{\bf 2}}
\newcommand{\bon}{{\bf 1}}

\def\ds#1{#1\kern-1ex\hbox{/}}
\def\sla{\raise.15ex\hbox{$/$}\kern-.57em}
\newcommand{\drawsquare}[2]{\hbox{%
\rule{#2pt}{#1pt}\hskip-#2pt%  left vertical
\rule{#1pt}{#2pt}\hskip-#1pt%  lower horizontal
\rule[#1pt]{#1pt}{#2pt}}\rule[#1pt]{#2pt}{#2pt}\hskip-#2pt%  upper horizontal
\rule{#2pt}{#1pt}}% right vertical
\newcommand{\fund}{\raisebox{-.5pt}{\drawsquare{6.5}{0.4}}}%  fund
\newcommand{\Ysymm}{\raisebox{-.5pt}{\drawsquare{6.5}{0.4}}\hskip-0.4pt%
         \raisebox{-.5pt}{\drawsquare{6.5}{0.4}}}%  symmetric second rank
\newcommand{\Yasymm}{\raisebox{-3.5pt}{\drawsquare{6.5}{0.4}}\hskip-6.9pt%
        \raisebox{3pt}{\drawsquare{6.5}{0.4}}}%  antisymmetric second rank

\begin{document}
%%%%%%%%%%%%%%%%%%%%%%%  FRONTESPIZIO  %%%%%%%%%%%%%%%%%%%%%%%%%%%%%%%%%%%
\begin{titlepage}
\begin{flushright}
{ROM2F/2010/11}\\
\end{flushright}
%%%%%%%%%%%%%%%%%%%%%%%  TITOLO  %%%%%%%%%%%%%%%%%%%%%%%%%%%%%%%%%%%%%%%%%
\begin{center}
{\large \sc Dynamical Supersymmetry Breaking in Intersecting Brane Models }\\
\vspace{1.0cm}
%%%%%%%%%%%%%%%%%%%%%%%%%  AUTORI  %%%%%%%%%%%%%%%%%%%%%%%%%%%%%%%%%%%%%%%%%
{\bf F.Fucito}, {\bf A. Lionetto}, {\bf J. F. Morales}, and   {\bf R. Richter}\\
 I.N.F.N. Sezione di Roma Tor Vergata\\ and \\{\sl Dipartimento di Fisica, Universit\'a di Roma ``Tor Vergata''\\
Via della Ricerca Scientifica, 00133 Roma, Italy}\\
\end{center}
\vskip 2.0cm
%%%%%%%%%%%%%%%%%%%%%%%%%  ABSTRACT  %%%%%%%%%%%%%%%%%%%%%%%%%%%%%%%%%%%%%%
\begin{center}
{\large \bf Abstract}\\\end{center}
 In this paper we study dynamical
supersymmetry breaking in absence of gravity  with the matter content of the
minimal supersymmetric standard model. 
The hidden sector of the theory is a strongly coupled gauge theory, realized in terms
of microscopic variables which condensate to form mesons. The supersymmetry breaking scalar potential
combines F, D terms with instanton generated interactions in the Higgs-mesons sector.
We show that for a large region in parameter space the vacuum breaks in addition 
to supersymmetry also electroweak gauge symmetry. We furthermore present local D-brane configurations that 
realize these supersymmetry breaking patterns.

\vfill
\end{titlepage}

\tableofcontents

\setcounter{section}{0}

\section{Introduction}

Breaking supersymmetry (SUSY) has always proved to be a
challenging and daunted task. Of all the possible options,
dynamical SUSY breaking (DSB) remains one of the most exciting
and economical choices of breaking SUSY. In this framework, SUSY
is a symmetry of the effective action and it is broken
spontaneously by the choice of the field theory vacuum. As it is
well known, the original toy models, in which SUSY was broken by
giving a vacuum expectation value (vev) to the F or D-terms, gave
a sparticle spectrum at odds with observations. This is why in the
usual SUSY extensions of the Standard Model (MSSM), the breaking is
achieved via the introduction of suitable terms in the Lagrangian
called soft SUSY breaking terms. They do not spoil the divergence
properties of the theory and also participate to the mechanism for the
gauge symmetry breaking of the theory. SUSY breaking is thus
intimately connected with all the main features of the MSSM. The
common lore wants soft SUSY breaking terms to be generated at
higher energies with respect to the mass of the sparticles in a so
called hidden sector. SUSY breaking is then communicated to the
visible sector via (gauge or gravity) messenger particles.
Recently many good books on SUSY have appeared in which these
issues are considered at length
\cite{Mohapatra,Drees,Terning:2006bq,Baer,Binetruy,Dine:2007zp}.

The progress in the computation of non perturbative effects in
gauge and SUSY theories, has opened up a new possibility for SUSY
breaking: the classical potential has a trivial vacuum which is
corrected by quantum effects. The quantum potential then exhibits
DSB. Already at the level of SUSY gauge theories, the existence of
condensates, driven by  non perturbative effects, does not
immediately translates into SUSY breaking. Each case must be
carefully examined: the standard criterion is to look for a global
symmetry, spontaneously broken, in  a theory which has no flat
directions. Another standard strategy is to look for
inconsistencies in the theory i.e. for a clash between two
different conditions which have to be simultaneously satisfied in order to preserve SUSY (for
example existence of a condensate which then violates the Konishi
anomaly)
\cite{Affleck:1984mf,Affleck:1984xz,Amati:1988ft,Shifman:1999mv}.
These results were first obtained for non chiral QCD like theories.
Their extension to chiral models of SUSY GUT type was
discussed in \cite{Amati:1988ft} and in the framework of gauge
mediation in \cite{Dine:1993yw,Dine:1994vc,Dine:1995ag}.

How to incorporate all of this into the framework of string theory
is then a completely different problem. Recently some of the
authors of the present paper \cite{Bianchi:2009bg} have analysed a
quiver model realizing a dynamical SUSY breaking scenario for a 
GUT theory. In this paper we want to take a further step and consider string intersecting
models with the matter content of the MSSM. There has been much work recently to embed
the SM and MSSM using intersecting branes:\footnote{
For recent reviews see  
\cite{Lust:2004ks,Uranga:2005wn,Blumenhagen:2005mu,Blumenhagen:2006ci,Marchesano:2007de}.}. 
here we discuss SUSY breaking in this scenario.
Therefore we consider a model where SUSY is broken
spontaneously via a non-perturbatively generated superpotential in
a hidden strongly coupled gauge theory.  The non-perturbative
superpotential will induce spontaneous breaking of gauge and SUSY
via an articulate conspiracy of gauge (D-terms) and Yukawa
(F-terms) interactions. Some of the steps we will take resemble
the so called KKLT approach followed in \cite{Kachru:2003aw} to stabilize some of
the moduli of a string theory compactification and more recently
in \cite{Dudas:2005vv,Dudas:2006gr,Dudas:2007nz}. However in our case
gravity plays no role: we just focus on the open string sector. We
thus postulate that the details of the compactification are
already taken care of, that moduli have been (completely or partially) stabilized and
that the closed string dynamics is not affecting our results.  
As we will see, these positions are not oversimplifying our model. The
open string sector must obey tight constraints to be consistent,
regardless of the closed strings. 

This is the plan of the paper: in Sections 2 and 3 we will discuss
our model from the point of view of field theory. In Sections 4 and 5
we will discuss local D-brane configurations which, at low
energy, lead to the field theories of the previous Sections.

\section{Susy-breaking extensions of the MSSM }

In this work we present two extensions of the MSSM which naturally give rise to SUSY and electroweak gauge symmetry breaking. In both cases the 
interplay of F and D-terms is crucial for the SUSY and electroweak gauge symmetry breaking. As we will see later in 
Section \ref{sec consistent string realizations} such extensions of the MSSM can naturally arise from D-brane compactifications.

Let us start by stating the gauge symmetry of the setups
\begin{align}
SU(3)_C\times SU(2)_L\times U(1)_Y\times U(1) \times  SU(3)_{H} \,\,,
\end{align}
which in addition to the usual SM gauge symmetry contains an abelian gauge symmetry and a strongly coupled hidden $SU(3)_H$.
%In the specific string embeddings extra $U(1)$ symmetries will be dressed this simplified setup. 

The chiral matter sector contains the usual MSSM matter content\footnote{Note that we include here also the right-handed neutrino $\nu^c$.}, 
namely the quarks and leptons
\be
   \vec \Phi = \{ Q_L, L, u^c, d^c, e^c, \nu^c\}\,\,,
\ee
which are collected in the vector $\vec\Phi$, where capital letters refer to left-handed superfields, while lower case letters denote right-handed fields. 
In addition to the MSSM particles we have the Higgs fields $H_{u,d}$ and two quark anti-quark pairs $Q_i$ and $\widetilde{Q}_i$ , $i=1,2$, charged with respect 
to the hidden $SU(3)_H$ and the additional $U(1)$. Moreover, depending on the considered setup there may be an additional field $Y$, which is neutral 
under the SM gauge groups and only charged under the $U(1)$.

The hidden $SU(3)_H$  gauge theory with  two quark anti-quark pairs $Q$ and $\widetilde{Q}$ will condensate  
via the generation of a Affleck, Dine and Seiberg 
superpotential
\be
W_{\rm non-pert}=\frac{\Lambda^7}{{\rm det} \,( M_{ij})}  \,\,,\label{wads}
\ee
where  $M_{ij}= Q_{i} \widetilde Q_{j}$ is the meson matrix. At the scale $\Lambda$ the gauge theory is effectively described in terms of the
mesons and baryons of $SU(3)_H$ which can be taken as the microscopic degrees of freedom of the low energy physics.
In the following we will denote by  $ \Lambda M_{i}$, the eigenvalues of the meson matrix $ M_{ij}$ and will work with the $M_i$ 
as fundamental degrees of freedom. Thus the SUSY - electroweak gauge symmetry breaking matter content, the Higgs-meson sector, is given by
\be
\vec X=\{ H_u,H_d, M_i,Y\}\,\,,
\ee
where the presence of $Y$ depends on the choice of the specific setup.

As we will see later in Section \ref{sec consistent string realizations} in D-brane compactifications the hypercharge is a linear combination of various $U(1)$'s. Explicitly
%From the string theoretical point of view it is more convenient to work in the string basis in which the hypercharge and
%the additional $U(1)$ take the form
\begin{align}
U(1)_Y =\frac{1}{2} U(1)_d+ \frac{1}{2} U(1)_e + ... \qquad U(1) =\frac{1}{2} U(1)_e- \frac{1}{2}U(1)_d \,\,,
\end{align}
where the dots indicate that there are further contributions for the hypercharge from other $U(1)$'s under which the Higgs-meson 
fields are uncharged. For later convenience  we label each $U(1)$ by a subscript that will later specify its D-brane origin.   
 
Below in Table \ref{Tabletoy} we display the non-trivial charges of the fields $\vec X$ in the Higgs-meson sector with
respect to the $SU(2)$ electroweak and $U(1)_{d,e}$ symmetries.
     \begin{table}[h]
  \centering
  \begin{tabular}[h]{|c|c|c|c|c|c|}
    \hline 
        & $H_u $ & $H_d $ &$M_1 $& $ M_2 $&Y  \\
    \hline  \hline
     $SU(2)$    & ${\bf 2} $ & ${\bf 2} $ &$ {\bf 1} $& $ {\bf 1} $& $ {\bf 1} $  \\  \hline
  %  $U(1)_Y$    & $\frac{1}{2}$  &$-\frac{1}{2}$ &$0$ & $0$ &$0$\\  \hline
     $U(1)_d$    & $0 $ & $-1 $ &$ 1 $& $ -1 $   & $1$\\  \hline
     $U(1)_e$    & $1 $ & $0 $ &$ -1 $& $ 1 $ & $-1$ \\  \hline
  \end{tabular}
   \caption{\small $SU(2)\times U(1)_d \times U(1)_e$-charges of the Higgs-meson sector}
   \label{Tabletoy}
\end{table} 
With that choice of charges the generic superpotential containing only the fields $\vec{X}$ and respecting the abelian symmetries $U(1)_d$ and $U(1)_e$ is of the form
\begin{align}
W= \mu H_u H_d  M_1 +m M_2  M_1 +\frac{\Lambda^5}{M_1 M_2} +\Big(\mu_Y  H_u H_d   Y +m_Y  M_2   Y \Big) \,\,.
\label{eq generic superpotential}
\end{align}
In absence of the field $Y$, clearly the last two terms of \eqref{eq generic superpotential} are absent.

In order to ensure that the color and electromagnetic gauge symmetries remain unbroken, we require the vanishing 
of all the vev's of the MSSM matter fields $\vec{\Phi}$. We look then for solutions with 
\begin{align}
\langle  \vec \Phi  \rangle = 0    \qquad       \langle  \vec X \rangle = x+\theta^2 F_x    \label{phizero}
\end{align}
In the following we will try to find SUSY and electroweak gauge symmetry breaking minima satisfying (\ref{phizero}). 
Thus we extremize the scalar potential only with respect to the fields $\vec{X}$. 
We will show in Section \ref{sec Susybreaking} that for a wide range in parameter space, SUSY and the electroweak gauge symmetries are broken 
in the $\vec{X}$ sector.

 \subsection{The Lagrangian}

After discussing the general setup above let us now describe in more details the Lagrangians of the models we will consider. 
For simplicity we take a canonical K\"ahler potential for the Higgs-meson fields $\vec X$. 
 More precisely we take the K\"ahler potential to be
\begin{align}
K(\vec{X},\vec{X}^{\dagger},\vec{\Phi}, \vec{\Phi}^{\dagger}) =\vec{X} \vec{X}^{\dagger} + 
 k_{ij}(\vec{X}, \vec{X}^{\dagger} ) \Phi_i  \Phi_j^{\dagger} + ...\,\, \,\,.
\label{eq Kahlerpotential}
\end{align}
Thus the K\"ahler metric for the scalar fields $\vec{X}$ will take a canonical form after taking into account the vanishing vev's of the 
MSSM matter fields $\vec{\Phi}$. We allow for a general $\vec X$-dependence of the functions $k_{ij}$ 
specifying the K\"ahler metric for the matter fields $\vec{\Phi}$.  When SUSY
is broken $F_X\neq 0$, these K\"ahler interactions provide soft symmetry breaking mass terms (of order $k_{ij}'' |F_X|^2$) for sparticles. 
These functions should then satisfy the phenomenological requirement that  the masses of sparticles are  beyond the observable limits. 
 Let us stress though that the specific type of K\"ahler potential does not affect our conclusions  regarding SUSY  and gauge symmetry breaking. 
% Any other K\"ahler potential whose form ensures a lift of squark and slepton masses compatible 
%with experimental observations is equally good.
 A similar $\vec X$-dependence can be introduced in the  gauge kinetic functions $\tau_a(\vec X)$ 
 in order to induce soft symmetry breaking masses (of order $\tau_a' F_X $) for the gauginos of the unbroken 
 symmetries $SU(3)_C\times U(1)_{\rm em}$ of  the  MSSM . 
  
With these assumptions the Lagrangians leading to the D and F terms can be written as
 \bea
&&  \L_{\rm D} = \left( k_{ij} (\vec X,\vec X^\dagger) \,  \Phi_i^\dag e^{  V  }  \Phi_j  +   \vec X^\dag e^{ V  }  \vec X + \xi_a V_a \right)
 \Big|_{\theta^2\bar \theta^2}  + {\rm h.c.}   \nn\\
&&  \L_{F} = \left[ \tau_a(\vec X)  \Tr (W^{(a)} W^{(a)}) +W(\vec X,\vec \Phi)\right]
 \Big|_{\theta^2}+ {\rm h.c.}\,\,.
\eea
 with $k_{ij}(\vec X, \vec X^\dagger)$, $\tau_a(\vec X)$ the quark and lepton K\"ahler function and $W(\vec X,\vec \Phi)$ the superpotential.
We will split the latter into two terms according to the number of quark and lepton superfields involved
\bea
W(\vec X,\vec \Phi)=W_0(\vec X)+W_2(\vec X,\vec \Phi)\,\,.
\eea
%We are interested in a SUSY breaking vacuum that preserves $SU(3)_C$  and the electromagnetic gauge symmetry. 
%This implies that all vev's of $\vec\Phi$ do vanish. 
$W_0$ determines the field theory vacuum.
In addition we would like to break SUSY and the electroweak gauge symmetry in 
such a way that $W_2( \langle \vec X \rangle ,\vec \Phi)$ gives realistic masses to the quarks and leptons.  This implies that 
the Higgs fields should acquire non-zero vev's.
Moreover, we  are interested in non-vanishing vev's for the $ F_{\vec X} $ in order to lift the masses of the sparticles 
compared to their SM partners.

 The pattern of SUSY and gauge symmetry breaking 
crucially depends on the choice of gauge 
and  Yukawa couplings, masses as well as on the Fayet-Iliopoulos terms entering the low energy action. 
In a string realization of this scenario, which will be discussed later, all these couplings and Fayet-Iliopoulos terms will be given by
the closed string background in which the D-brane setup is localized and will be input parameters in our analysis.
 We stress the fact that once closed string dynamics is turned on, what we refer as Fayet-Iliopoulos terms here become field 
dependent functions of the charged closed string moduli.

 \section{The Higgs-meson sector
\label{sec Susybreaking}
  }

In this Section we study the vacuum structure of the field theory models for two simple choices of $W_0$ of the type displayed in 
equation \eqref{eq generic superpotential}. In both cases, given reasonable choices for the parameters of the theories, a vacuum can be 
found which breaks SUSY and the electroweak gauge symmetry. This breaking requires an interplay between the F- and D-terms.
We start by analysing a configuration which exhibits the fields $H_u$, $H_d$, $M_1$ and $M_2$ in the Higgs-meson sector.
Later we  will allow for an additional field $Y$, charged under $U(1)_d$ and $U(1)_e$.

%The form of these superpotentials, their associated field content and their couplings to the MSSM particles  will be derived
%in the next Section from quiver gauge theories living at type IIA brane intersections.

  \subsection{First Example
  \label{sec D-term breaking}}

Let us consider the following  field content in the Higgs-meson sector
 \be
 \vec X=\{ H_u, H_d, M_1, M_2\}  \,\,.
 \ee
The charges of the various fields were displayed in Table \ref{Tabletoy}.
 For the superpotential we take
 \be
 W_0=\mu H_u H_d M_1+ m M_1 M_2 +\frac{\Lambda^5}{M_1 M_2} \label{w0}\,\, ,
 \ee
where the latter term is the non-perturbative ADS superpotential of the hidden $SU(3)_H$.
In order to preserve the electromagnetic gauge symmetry  we look for vacuum solutions of the form
    \begin{align}
  & \hspace{0mm}H_u =
\left(
\begin{array}{c}
  0  \\
  h_{u}+\theta^2 F_u    
\end{array}
\right)  \qquad H_d=
\left(
\begin{array}{c}
  h_{d}+\theta^2 F_d  \\ \label{eq vevs}
  0   
\end{array}
\right)  \nn\\ \\ \nn
 M_i &= 
  x_i +\theta^2 F_{x_i}    \qquad 
V_a  =  \theta^2  \bar\theta^2 D_a  \qquad W^{\a}_a=\theta^\a D_a  \,\,
  \end{align}
  and take $\tau_{a}={1\over g^2_{a}}$ for the $a=d,e,SU(2)$ components. 
In terms of these variables the scalar potential can be written as
    \be
  V=|\vec F |^2 +\frac{1}{2 g_a^2}\, \vec D_a^2\,\,.
  \ee 
Here the F-terms take the form
\begin{align}
\bar F_u = \mu\,  x_1 h_d \qquad  \bar F_d &= \mu\,  x_1 h_u  \qquad \bar F_{x_2} =-m x_1+\frac{\Lambda^5}{x_1  x_2^2}  \nn \\
 F_{x_1} &= \mu\,   h_u h_d -m x_2  +\frac{\Lambda^5}{x_1^2  x_2}\,\,,
 \end{align}
whereas the $D$ terms are given by
\begin{align} \nn
  &D_{SU(2)}  =  \frac{g^2_{SU(2)}}{2}(|h_d|^2-|h_u|^2  )  \qquad D_{d} =  g^2_{d} \left(  - |h_d|^2+ |x_1|^2 -|x_2|^2+\xi_d \right) \\ 
  & \hspace{3cm }D_{e} =  g^2_{e} \left(   |h_u|^2- |x_1|^2+  |x_2|^2 +\xi_e    \right)\,\,.
\label{formfi}\end{align}
In (\ref{formfi}) we included a Fayet-Iliopoulos term for the $U(1)$'s. We will later discuss under what circumstances SUSY and 
the electroweak gauge symmetry are broken.

  \subsubsection*{Supersymmetric solution}
  
Let us first consider the case in which SUSY is unbroken. Such solutions can be found for
 $\xi_{d}+\xi_{e}=0$ and are given by
  \be
  h_u=h_d=0 \qquad x_1 \cdot x_2=\left( \frac{\Lambda^5}{m}\right)^{1\over 2} \qquad |x_1|^2-|x_2|^2=\xi_e\,\,.  \label{solsusy1}
  \ee
For vanishing FI-terms $\xi_{d}=\xi_{e}=0$ the solution takes the simple form 
  \be
  h_u=h_d=0 \qquad x_1 = x_2=\left( \frac{\Lambda^5}{m}\right)^{1\over 4}\,\,.  \label{solsusy2}
  \ee 
  Note that these supersymmetric solutions do not break the electroweak gauge symmetry. 
  
   \subsubsection*{Non supersymmetric solution with gauge symmetry unbroken }

In the following we analyze the effect of generic Fayet-Iliopoulos terms. We start by looking for a vacuum in which 
the $SU(2)$ gauge symmetry is unbroken, i.e. $h_u=h_d=0$. 
In this case the equation of motion can be easily solved by taking all the fields to be real  and
  \bea
%  h_u=h_d=0,\, 
x_1= \left(\frac{\Lambda ^5}{m}\right)^{\frac{1}{4}}\sqrt{\sqrt{\Delta ^2+1}-\Delta} 
   \qquad    
   x_2= \left(\frac{\Lambda^5}{m}\right)^{\frac{1}{4}}\frac{1}{\sqrt{\sqrt{\Delta ^2+1}-\Delta} } \label{almostsusy}
  \eea
  with
  \be
  \Delta^2= {  m(g_d^2\, \xi_d -g_e^2\, \xi_e )^2 \over 4 \Lambda^5 (g_d^2+g_e^2)^2} \,\,.
  \ee
   At the minimum $ F_{\vec X}=0$ and the potential   takes the form
   \be
   V={1\over 2} \frac{(\xi_d+\xi_e)^2g_d^2 g_e^2 }{g_d^2 +g_e^2}\,.
   \ee 
For $\xi_d+\xi_e = 0$ the scalar potential is vanishing indicating a SUSY solution. However for
 $\xi_d+\xi_e\neq 0$ SUSY is broken. It is instructive to look at the simplest example of
a non-SUSY solution in this class $\Delta=0$, i.e. for couplings satisfying
    \be
    \xi_d g_d^2-\xi_e g_e^2=0\,\,.
    \ee
     For this choice the linear terms in $\xi_{d,e}$ in the D-term scalar potential cancel each other
     at $h_{u,d}=0$ and one finds
       \begin{align*}
    \frac{1}{g^2_d} \,D_d^2+  \frac{1}{g^2_e} \, D_e^2= {1\over 2} \frac{(\xi_d+\xi_e)^2g_d^2 g_e^2 }{g_d^2 +g_e^2}+ {1\over 2}(g^2_{e}+g_d^2) \left( |x_1|^2- |x_2|^2\right)^2+ {\cal O} (h^2)
     \end{align*}
     with an extremum at (\ref{solsusy1})\footnote{Notice that for $\Delta=0$ the non-supersymmetric solution (\ref{almostsusy}) 
     reduces to (\ref{solsusy1}).} . 
     This is a minimum if the masses of $h_{u,d}$ at the extremum are positive. Expanding
     to second order in $h_{u,d}$ one finds
     \begin{align*}
     V\big|_{ext}=  {1\over 2} \frac{(\xi_d+\xi_e)^2g_d^2 g_e^2 }{g_d^2 +g_e^2}+|h_u|^2 (\mu^2 |x_1|^2+g_e^2 \xi_e)\Big|_{ext} +
|h_d|^2 (\mu^2 |x_1|^2-g_d^2 \xi_d)\Big|_{ext} + {\cal O} (h^4)\,\,.
     \end{align*}
Taking FI terms positive, one finds a positive mass for $h_d$ if  
         \be
   \Lambda^5 > {m \,  g_d^4\, \xi^2_d\over \mu^4} \label{cond}
\ee     
     We conclude that for  $\xi_d+\xi_e \neq 0$ SUSY is broken if (\ref{cond}) is satisfied.
 
     Alternatively one can explicitly compute the eigenvalues of the Hessian matrix
  and finds 
  \bea
  \partial_{ij} V &=& {\rm diag}  \left(   8(g_d^2+g_e^2)  \left({\Lambda^5\over m}\right)^{1\over 2} ,32 (\Lambda^5 m^3)^{1\over 2} ,32 m^2,\right. \nn\\
 &&\left. 2\mu^2   \left({\Lambda^5\over m}\right)^{1\over 2} + 2 g_d^2 \xi_d ,  2\mu^2   \left({\Lambda^5\over m}\right)^{1\over 2} - 2 g_d^2 \xi_d,0,0,0
   \right)\,\,.
  \eea
  The last three entries in this matrix are the goldstone bosons associated to the breaking of the three $U(1)$ symmetries (with the third    
  U(1) coming from the Cartan of $SU(2)$).  The remaining eigenvalues are then positive if (\ref{cond}) is satisfied.

 Let us display an explicit solution for a concrete choice of parameters 
     \begin{align*}
&\hspace{5mm} g_{SU(2)}=g_{d}=g_{e}=0.5 \qquad \Lambda=m=\mu =\xi_e=\xi_d=1\\\\
& \hspace{5mm} h_u=h_d=0 \qquad x_1=x_2=1  \qquad V=0.25\,\,.
\end{align*}
Here the dimensionful quantities $m$, $\Lambda$  as well as the vev's of the fields are given in units of, let us say, TeV, whereas the 
FI-terms $\xi_d$, $\xi_e$ are measured in units of ${\rm TeV}^2$ and $V$ in ${\rm TeV}^4$.

 \subsubsection*{Non supersymmetric solution with gauge symmetry broken}     
 Let us now turn to the second type of solutions in which not only SUSY is broken but also the electroweak gauge symmetry is.
For simplicity we restrict ourselves to the case 
\be
  \xi_d g_d^2=\xi_e g_e^2
\ee
which breaks gauge and SUSY if 
 \be
0< \Lambda^5 < {m \,  g_d^4\, \xi^2_d\over \mu^4}\,\,.
\ee
It is hard to find analytic solutions but a numerical analysis is viable and shows that such configurations lead to a minimum with
$h_u,h_d \neq 0$. A representative of the latter is given by
\begin{align*}
&\hspace{5mm} g_{SU(2)}=g_{d}=g_{e}=\Lambda=0.5 \qquad m=\mu =\xi_e=\xi_d=1\\\\
& \hspace{5mm} h_u=0.01  \qquad h_d=0.5 \qquad x_1=0.37  \qquad x_2=0.46 \qquad V=0.23\,\,,
\end{align*}
where again the dimensionful quantities $m$, $\Lambda, \xi_{d,e}$  as well as the vev's of the fields are given 
in the same units of our previous solutions.
In this solution, the Higgs fields acquire a non-zero vev triggering the breaking of the electroweak gauge symmetry.

\subsection{Second Example
\label{sec f-term breaking}}

Now we allow for an additional field $Y$. Thus we consider the following  ${\cal N}=1$ chiral field content in the Higgs-meson sector
 \be
 \vec X=\{ H_u,H_d,M_1,M_2, Y\}  \,\,.
 \ee
 In Table \ref{Tabletoy} we display  the charges of the various fields . 
 The superpotential containing the fields $\vec{X}$ and obeying the gauge symmetries is now given by    
 \be
 W_0=  \mu\, M_1 H_u H_d+\mu_Y  \,Y  H_u H_d  +m\,M_2 M_1 + m_Y\, Y M_2  +{\Lambda^5\over M_1 M_2}   \label{w02}\,\,,
 \ee
where the last term is due to the non-perturbative ADS superpotential generated in the hidden $SU(3)_H$ gauge theory.

In order to preserve $SU(3)_C$ and the electromagnetic gauge symmetry we again look for solutions of the type \eqref{eq vevs}. 
In addition we parametrize the field $Y$ as follows $Y=y+\theta^2 F_y$.
Then the F-terms take the form
  \begin{align}
  \bar F_{u,d} &=   h_{d,u}   (\mu x_1+\mu_Y  y )   \qquad  \hspace{7.5mm}\bar F_Y =   \mu_Y h_u h_d - m_Y x_2   
  \nn\\
    \bar F_{x_1} &=\mu \,   h_u h_d -m x_2  + {\Lambda^5\over x_1^2  x_2}\qquad   
    \bar F_{x_2} =-m x_1-m_Y y+{\Lambda^5\over x_1  x_2^2} 
  \end{align}
  and the D-terms are given by
  \begin{align}  
D_{SU(2)}  &=  \frac{g^2_{SU(2)}}{2}\Big(|h_d|^2-|h_u|^2  \Big) \qquad  D_{d} =  g^2_{d} \left(  - |h_d|^2+ |x_1|^2-  |x_2|^2+|y|^2 + \xi_d\right)   \nn \\
      & \hspace{20mm}   D_e =  g^2_{e} \left(   |h_u|^2- |x_1|^2+  |x_2|^2-|y|^2 + \xi_e\right) \,\,.
  \end{align}

In the following we analyze for which values of the parameters, SUSY and the $SU(2)_L \times U(1)_Y$ gauge symmetry are broken.
  \subsubsection*{Supersymmetric solution}
Before discussing the broken phase  let us again first look for supersymmetric solutions. Such a solution exists if the parameters satisfy
  \be
  \mu_Y\neq 0 \qquad {\rm and} \qquad \Delta =m \mu_Y-\mu m_Y \neq 0 \qquad  \xi_d=\xi_e=0\,\,.
  \ee
  The supersymmetric solution can in that case be written as
  \bea
y=-{\mu x_1\over \mu_Y} \qquad   h_u=h_d=\left( { \Lambda^5 m_Y^2   \over x_1^2 \mu_Y \Delta   }\right)^{1\over 4}
 \qquad   x_2=-\left( {\Lambda^5 \mu_Y \over x_1^2 \Delta }\right)^{1\over 2}   \label{soly}
  \eea
  with $x_1$ determined by the equation
  \be
  |x_1|^2-  |x_2|^2+|y|^2 =0
    \ee
    evaluated at (\ref{soly}). 
Let us point out that for these SUSY solutions the vev's of the Higgs are non-vanishing and the $SU(2)_L \times U(1)_Y$ symmetry is broken.

 \subsubsection*{Non-supersymmetric vacua}
   
Once again switching on Fayet-Iliopoulos terms gives SUSY
and electroweak gauge symmetry breaking terms for quite general choices of the parameters.  
An analytic solution for the non SUSY minimum is hard to find but the
 equations can be easily solved numerically for any choice of the gauge and Yukawa couplings. Let us display one example for 
each type of solution, one for unbroken electroweak gauge symmetry and one for the broken electroweak gauge symmetry.

\begin{itemize}

\item{  Gauge symmetry unbroken:
\begin{align*}
%&\hspace{45mm} \text{gauge unbroken} \\
&\hspace{5mm} g_{SU(2)}=g_{d}=g_{e}=0.5 \qquad \Lambda=m=\mu=m_y=\mu_y  =\xi_e=\xi_d=1\\
& \hspace{5mm} h_u=h_d=0 \qquad x_1=0.16 \qquad x_2=1.13 \qquad y=0.8   \qquad V=1.03
\end{align*}
  }
   \item{ Gauge symmetry broken
  \begin{align*}
% &\hspace{47mm} \text{gauge broken} \\
&\hspace{5mm} g_{SU(2)}=g_{d}=g_{e}=0.5 \qquad \Lambda=m=\mu=m_y=-\mu_y  =\xi_e=\xi_d=1\\
& \hspace{5mm} h_u=0.6 \qquad h_d=1.3 \qquad x_1=y=0.9  \qquad x_2=0.78   \qquad V=0.08
\end{align*}
}
\end{itemize}
where the dimensionful quantities $m$, $m_Y$, $\Lambda$ as well as the vevs of the vacuum solution are measured in units of TeV and the FI-terms 
are measured in units of ${\rm TeV}^2$.

 The vev's of $h_u$ and $h_d$ are different from zero only for the second solution which is thus breaking the
electroweak gauge symmetry.

\section{D-brane model building
\label{sec D-brane model building}}

In Section \ref{sec consistent string realizations} we will present some D-brane quivers which mimic the configurations discussed above. 
Before presenting these quivers let us briefly discuss, 
following \cite{Cvetic:2009yh} (see also \cite{Bianchi:2000de,Dijkstra:2004cc})\footnote{For analogous work see \cite{Ibanez:2008my,Leontaris:2009ci,Anastasopoulos:2009mr,Kiritsis:2009sf,Cvetic:2009ez,Cvetic:2010mm,Anastasopoulos:2010ca, Cvetic:2010dz,Blumenhagen:2010dt}. First local (bottom-up) constructions were discussed in \cite{Antoniadis:2000ena,Aldazabal:2000sa,Antoniadis:2001np}.}, the various constraints on the transformation 
properties of the chiral matter fields which arise from string theory and that not always have an analogue in the field theory context.

For concreteness we focus on type IIA  
string theory in which the basic building blocks are D6-branes which fill the four dimensional space-time and wrap a three-cycle in the internal 
compactification manifold. The gauge symmetry living on the worldvolume of a stack of $N$ D6-branes transforms under a $U(N)$  group.
%, which further splits  into an $SU(N) \times U(1)$, where the abelian $U(1)$ generically becomes massive via the Green-Schwarz mechanism.
 Chiral matter appears  at the intersection between two stacks of D6 branes, $a$ and $b$, and transforms as a bifundamental representation under $U(N_a)\times U(N_b)$. 

In order to obtain ${\cal N}=1$ in the four-dimensional spacetime one introduces an orientifold action, which implies the presence of 
$O6$-planes which fill out the spacetime and wrap orientifold invariant three-cycles in the internal manifold. Their
presence is crucial to cancel the RR-charge carried by the D6-branes. Moreover due to the presence of an orientifold action each stack of D6-branes 
has an orientifold image, which allows for chiral matter transforming in the symmetric or antisymmetric of the gauge group $U(N_a)$.
The multiplicities of the chiral matter fields are given in terms of the intersection numbers of the three-cycle $\pi_a$, $\pi_a'$ 
and $\pi_{O6}$, where $\pi_a$ is the 
three-cycle wrapped by the stack $a$, $\pi'_a$ its orientifold image and  $\pi_{O6}$ is the whole class of orientifold 
invariant three-cycles in the internal manifold.

The field content of an  intersecting brane model is determined by the intersection numbers 
according to Table  \ref{table chiral spectrum}. Here we denote by $\Delta n (\mathbf{R})$ the net number of chiral fields in a representation
$\mathbf{R}$, i.e.
\be
\Delta n (\mathbf{R}) = n(\mathbf{R})-
n (\mathbf{\bar R} )\,\,.
\ee
Note that although the antisymmetric representation of $U(1)$ does not exist the corresponding 
intersection number in the Table \ref{table chiral spectrum} may be not vanishing.  In the following,
 we will refer to this intersection number as $\Delta n (\Yasymm_a)$ 
even for $U(1)_a$ gauge groups where $\Yasymm_a$ does not exist. This intersection 
number will enter in the consistency conditions constraining the string model. 
This notation will then allow us to write the various string consistency conditions in a unifying 
compact way independently of the gauge group rank.  
   
    \begin{table}
\centering
\begin{tabular}{|c|c|}
\hline
Representation  & Multiplicity $\Delta n (\mathbf{R})$ \\
\hline $ \Yasymm_a$
%$[{\bf A_a}]_{L}$
 & ${1\over 2}\left(\pi_a\circ \pi'_a+\pi_a \circ  \pi_{{\rm O}6} \right)$  \\
$\Ysymm_a$
%$[{\bf S_a}]_{L}$
     & ${1\over 2}\left(\pi_a\circ \pi'_a-\pi_a \circ  \pi_{{\rm O}6} \right)$   \\
$( \fund_a,{\overline \fund}_b)$
%$[{\bf (\o N_a,N_b)}]_{L}$
 & $\pi_a\circ \pi_{b}$   \\
 $(\fund_a, \fund_b)$
%$[{\bf (N_a, N_b)}]_{L}$
 & $\pi_a\circ \pi'_{b}$
\\
\hline
\end{tabular}
\vspace{2mm} \caption{Chiral spectrum for intersecting D6-branes.}
\label{table chiral spectrum}
\end{table}

 Even in a local set up the tadpole cancellation condition constrains the spectrum of the string model. 
%In fact it implies the vanishing of non-abelian anomalies while
%abelian and mixed anomalies are cancelled via the Green-Schwarz mechanism. 
While for non-abelian gauge groups these constraints boil down to the usual anomaly cancellation conditions in field theory, in
presence of $U(1)$ symmetries they further constraint the spectrum of $U(1)$ charges in the string model. 

Generically, anomalous $U(1)$ acquire a mass via the Green-Schwarz
mechanism of anomaly cancellation.
Non-anomalous $U(1)$ gauge bosons can also become massive via non-trivial Chern-Simons (CS)
couplings with RR fields.  The massive $U(1)$'s are generically not part of the low
energy effective gauge symmetry but remain as unbroken global symmetries  at
the perturbative level and thus may forbid  various desired couplings. 
Since the standard model gauge symmetry contains the abelian subgroup $U(1)_Y$, we require that a
linear combination
\begin{align}
U(1)_Y= \sum_{x} q_x \,U(1)_x \,\,,\label{eq hypercharge}
\end{align}
remains massless. This happens if the CS coupling $\int_{D6} C \wedge F_{U(1)}$  vanishes for all the D6 branes.
The condition for the presence of a massless $U(1)_Y$ translates then  into a condition on the cycles wrapped by the D-branes, which together
with the tadpole cancellation condition constrains the charges  of the chiral matter field content. 
In the following we will discuss both  constraints in more detail.

\subsection*{Tadpole condition \label{sec tadpole}} 
The tadpole
condition is a constraint on the cycles wrapped by the D-branes. In the framework of intersecting D6-brane models the tadpole conditions read
\begin{align*}
\sum_{x} N_x \left(\pi_x +\pi'_{x} \right) -4\pi_{O6}=0\,\,.
\end{align*}
 Here $x$ denotes the different D-brane stacks present
in the model.  This condition ensures that the total RR charge carried by the D6-branes (and their images) exactly cancels that of the O6 planes.
Multiplying this equation with the homology class of the cycle
that is wrapped by a stack $a$ gives, after a few manipulations and using 
 the relations displayed in Table \ref{table chiral spectrum}
\begin{align}
\Delta n(\,\fund_a) + (N_a-4)\Delta n(\, \Yasymm_a) + (N_a+4) \Delta n
(\,\Ysymm_a)=0 \,\,,\label{eq constraint1}
\end{align}
where   
\be
\Delta n(\,\fund_a)= \sum_x N_x \Big[\Delta n (\fund_a,{\overline \fund_x}) + \Delta n (\fund_a,\fund_x) \Big] 
\ee
is the total number of $U(N_a)$ fundamentals (minus antifundamentals). 
Equation \eqref{eq constraint1} is
nothing else than the anomaly cancellation for non-abelian gauge
theories for $SU(N_a)$ gauge groups with rank $N_a>2$.  
We stress that also for $N_a=1,2$ equation \eqref{eq constraint1} leads to binding constraints with a less obvious field
 theory analogue.

\subsubsection*{Massless U(1)'s
\label{sec massless U(1)}}

Since the SM contains the $U(1)_Y$ hypercharge as a
gauge symmetry we require the presence of a massless $U(1)$ which can be identified with $U(1)_Y$.
Only for specific choices of the coefficients in (\ref{eq hypercharge})
the matter particles have the proper hypercharge.
 In addition, once the coefficients $q_x$ are given, the corresponding linear combination of the $U(1)'s$  
remains massless if the condition \cite{Aldazabal:2000dg}
\begin{align}
\sum_{x} q_{x}\,N_{x} (\pi_{x} -\pi'_{x}) =0
\label{eq cond for masllessness}
\end{align}
is satisfied. This condition ensures that the CS coupling $\int C \wedge F_{U(1)_Y}$ cancels.

Analogously to the analysis performed for the tadpole constraints we multiply
 \eqref{eq cond for masllessness} with the homology class $\pi_a$  of the
cycle wrapped by the D-brane stack $a$. After reinterpreting the intersection numbers in
terms of the multiplicities according to the Table \ref{table chiral spectrum} we obtain
\begin{align}
 \sum_{x \neq a} q_x\,
N_x \,  \left[\Delta n (\fund_a,{\overline \fund_x}) - \Delta n (\fund_a,\fund_x) \right]=q_a\,N_a \,\Big(\Delta n(\Ysymm_a) + \Delta n (\Yasymm_a)\Big) \,\,.
\label{eq massless constraint non-abelian}
\end{align}
We remark that this equation gives a constraint for every D-brane stack present in the model.
Thus for the six-stack model (which we will consider later) we expect six additional constraints due to the presence of a massless hypercharge.
   
  \section{Consistent string realizations
  \label{sec consistent string realizations} } 
Now we have all the ingredients to engineer some consistent string models based on intersecting  D6 branes and O6 planes for which the   field content contains  the MSSM particles and the  dynamical SUSY breaking Higgs-meson sector of the types we discussed in Section \ref{sec Susybreaking}.
We present string realizations based on six stacks of D6-branes which lead to the gauge symmetry
\be
G=U(3)_a\times SU(2)_b\times U(1)_c\times U(1)_d\times U(1)_e\times U(3)_f\,\,.
\ee
The stack $b$ is on top of an O6 plane, i.e.  $b=b'$, thus leading to a $Sp(1)\sim SU(2)$ gauge group. 
The hypercharge is of the form \cite{Ibanez:2001nd}
   \be
U(1)_Y=\sum q_x U(1)_x=\frac{1}{6} U(1)_a + \frac{1}{2} \left[ U(1)_c+U(1)_d+U(1)_e+
   U(1)_f \right] \,\,.
\ee
We consider models with one and three generations of MSSM particles and a Higgs-meson sector of the form discussed 
in Section  \ref{sec Susybreaking} which satisfies the tadpole and massless $U(1)_Y$ conditions. The Higgs-meson superpotential 
is of the types (\ref{w0}) and (\ref{w02}), respectively. As explained in Section \ref{sec Susybreaking} these configurations 
break SUSY and the electroweak gauge symmetry after condensation of the hidden $SU(3)_H$ gauge theory. The D-brane quivers have 
the same Yukawa interactions of the MSSM and satisfy some basic phenomenological requirements. The latter include the absence of R-parity 
violating couplings and of some dimension five operators which could lead to a disastrous short proton lifetime.
 
The field content of the discussed models is summarized in Table \ref{table allmodels}. Here $A=a, b, c, \ldots$  denotes the various brane stacks.
Their images with respect to the O6 orientifold plane are denoted by $A'=a', b', c' \ldots$.  The first column of the table displays the string 
origin of each state with $AB$  labelling an open string originating from the brane stack $A$ and ending on the brane stack $B$.     
Different models will be distinguished by different choices of the multiplicities $n$'s in the last column of the table.

   \begin{table}[h]
  \centering
  \begin{tabular}[h]{|c|c|c|c|c|c|c|c|c|}
    \hline  sector  & matter   & $U(3)_a$ & $SU(2)_b$ &$U(1)_c$& $U(1)_d$ & $U(1)_e$   & $U(1)_Y$& number \\
    \hline ab  &$Q_L$   & $\bth$       &  $ \btw$       &  $0$   & $0$& $ 0$  & $\frac{1}{6}$  & $n$\\
     \hline e'a &$u^c $   & $\bar \bth$  &  $\bon$       &  $0$  & $0$& $-1$   & $-\frac{2}{3}$  & $n_{u}$  \\
        \hline d'a &$u^c$   & $\bar \bth$  &  $\bon$       &  $0$  & $-1$& $0$   & $-\frac{2}{3}$  &$n-n_{u}$  \\
         \hline da &$d^c $   & $\bar \bth$  &  $\bon$       &  $0$ & $1$  & $0$ & $\frac{1}{3}$   & $n_{d}$\\ 
           \hline ea &$d^c$   & $\bar \bth$  &  $\bon$       &  $0$ & $0$  & $1$ & $\frac{1}{3}$   &$n-n_{d}$\\ 
    \hline bc &$L$   & $\bon$       &  $ \btw$       &  $-1$  & $0$ & $0$    & $-\frac{1}{2}$ & $n$\\
     \hline cd' &$e^c$ & $\bon$       &  $\bon$       &  $ 1$  & $1$ &$0$  & $1$  & $n-n_{e}$\\
     \hline ce' &$e^c$ & $\bon$       &  $\bon$       &  $ 1$  & $0$ &$1$   & $1$  &$n_{e}$\\
       \hline ce &$\nu^c$ & $\bon$       &  $\bon$       &  $ 1$  & $ 0$ &$-1 $  & $0$  &$n- n_{\nu}$ \\
       \hline cd &$\nu^c$ & $\bon$       &  $\bon$       &  $ 1$  & $ -1$ &$0 $  & $0$  &$ n_{\nu}$ \\
    \hline bd &$H_d$ & $\bon$       &  $ \btw$       &  $0$   & $-1$&$0$   & $-\frac{1}{2}$  &1\\
    \hline e'b &$H_u$ & $\bon$       &  $ \btw$       &  $0$  & $0$ &$1$   & $\frac{1}{2}$  &1\\ 
        \hline
         \hline   de & $Y$  & $\bon$       &  $ \bon$       &  $0$  & $1$ &$-1$   & $0$  & $n_Y$ \\ 
             \hline  e'ffd' &$M_1 $   & $\bon$       &  $\bon$       &  $0$     & $1$& $-1$   & $0$  &1\\
              \hline  d'ffe'  &$M_2 $   & $\bon$       &  $\bon$       &  $0$     & $-1$& $1$    & $0$  &1\\
         \hline
  \end{tabular}
   \caption{\small Models with SUSY breaking. The multiplicities in the last column depend on the 
   specific model. $n$ is the number of generations and $n_Y=1,0$ for models with or without Y, respectively.
   The remaining multiplicities $(n,n_u,n_d,n_e,n_{\nu})$  are $(1,1,1,1,1)$ or $(3,2,2,2,2)$ for the models without the Y field
and $(1,1,0,0,0)$ or $(3,2,2,2,3)$ for the models with a Y field  and one or three generations respectively.}
   \label{table allmodels}
\end{table}
  The field content of the various  models can be alternatively summarized by gauge quiver diagrams. In the latter,
each node stands for a D brane stack carrying a definite gauge groups and each arrow for an open string connecting
two stacks and thus giving a chiral matter multiplet. 
Two arrows connecting two different brane stacks stand for a chiral matter multiplet transforming in the bifundamental representation
of the gauge groups, while those arrows connecting a brane stack and its
image account for chiral matter multiplet transforming in the symmetric, or antisymmetric, representation
of the gauge groups. For simplicity we will present only string models with no symmetric or antisymmetric matter but more general choices
with a similar phenomenology are allowed.
    
The tadpole condition translates into the requirement that the number of
arrows arriving at a node is equal to the number of arrows leaving it. To perform this counting, the flavour multiplicity
of each arrow must be kept into account. 
For an arrow entering a certain node, its flavour group is given by the number of branes in the stack at the opposite end of the arrow.
Finally image nodes contribute with an opposite sign while an extra $\pm 4$ orientifold contribution should be taken into account
for an arrow connecting a node to its image. 
The massless $U(1)_Y$-condition \eqref{eq cond for masllessness} leads to the same counting weighted by the $U(1)_Y$ charge $q_x$ of the
flavor node $x$. In this case there is no extra contributions from the arrows crossing the O6 plane since the O6 plane does not contribute to the
CS coupling.

 \subsection{First Example} 

In Figures \ref{fig dquark1} and \ref{fig dquark2} we display a one and three generation configuration which exhibits a SUSY breaking superpotential for the 
Higgs-meson sector of the type \eqref{w0} discussed in Section \ref{sec D-term breaking}.  

\subsubsection*{One generation quiver}   
   
Let us start by analysing a one generation quiver which mimics the  SUSY breaking configuration discussed in Section \ref{sec D-term breaking}.  
   \begin{figure}[h]
\begin{center}
 \includegraphics[width=0.9\textwidth]{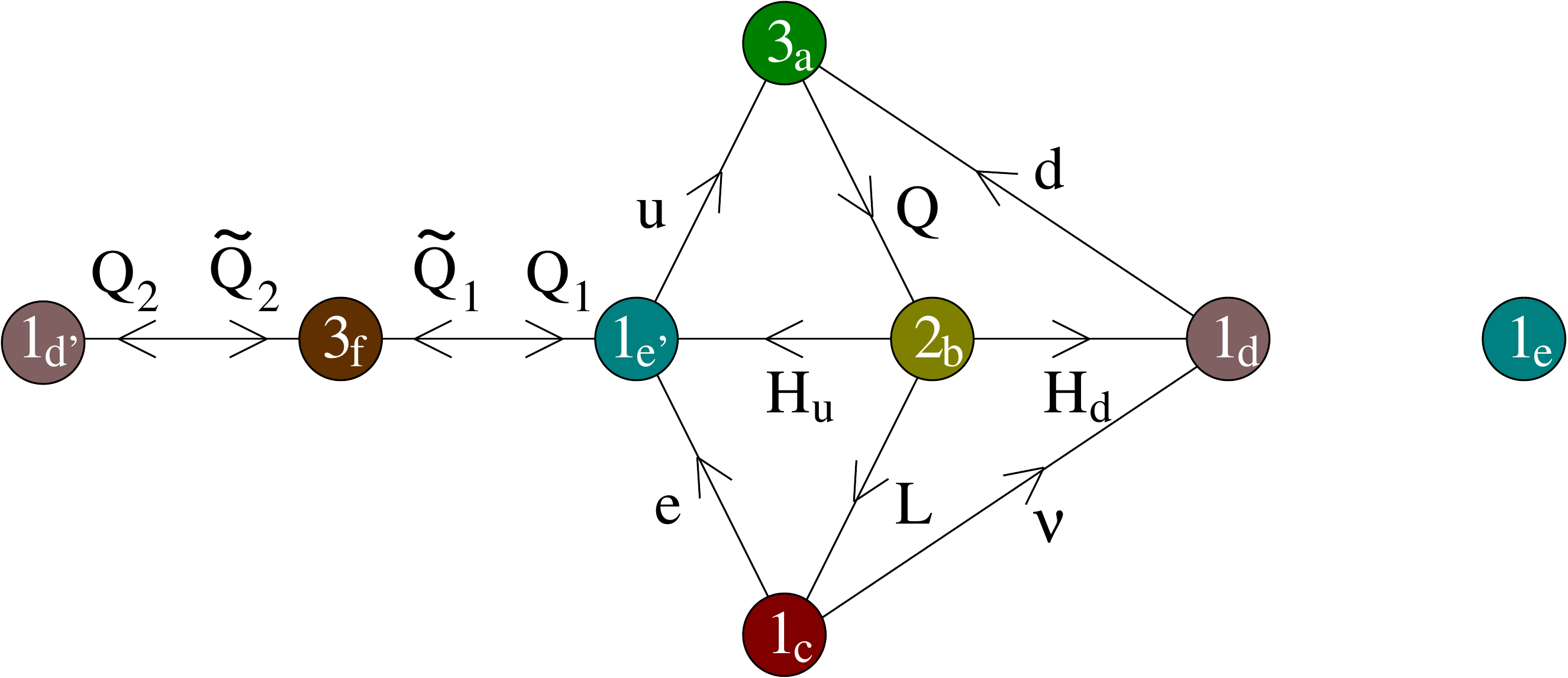}
\end{center}
\caption{\small One generation quiver leading to SUSY breaking. This diagram is symmetric with respect to the orientifold O6 plane 
located on the b stack of branes. Certain mirror nodes are omitted for the sake of simplicity.}\label{fig dquark1}
\end{figure}
The choice of the multiplicity numbers in Table \ref{table allmodels} is 
\begin{align*}
n=1\qquad n_u=1 \qquad n_d=1 \qquad n_e=1 \qquad n_{\nu}=1\qquad n_Y=0\,\,,
 \end{align*}
which satisfies the constraints arising from tadpole cancellation as well as the masslessness of the hypercharge $U(1)_Y$ discussed in Section \ref{sec D-brane model building}. That choice leads for the MSSM spectrum to
\begin{align*}
 Q_L&=(a,b) \qquad u^c=(\bar a, \bar e) \qquad d^c=(\bar a,  d) \\
  L&=(b , \bar c) \qquad \hspace{1mm} e^c =(c,e) \qquad \nu^c=(c, \bar d)
 \end{align*}
and for the Higgs-meson sector to
 \begin{align}
 H_u=(b,e) \qquad H_d=(b,\bar d) \qquad M_1 =(d, \bar e)\qquad M_2 =   (\bar d, e) \,\,.
 \label{eq Higgs meson sector D-term}
\end{align}
Here the mesons arise after condensation and will be given in terms of $Q$ and $\widetilde{Q}$
 \begin{align}
 M_1 =   \widetilde{Q}_1 \,Q_2   \qquad \text{and}\qquad   M_2 = \widetilde{Q}_2 \, Q_1 \,\,.
 \end{align}

The perturbative superpotential is given by all closed loops in the quiver diagram, where one can jump
from a node $x$ to its orientifold image  
changing the orientation of the loop. 
Let us perform the analysis concretely for the superpotential in the Higgs-meson sector, which is given by 
%It is easy to check that the quiver indeed exhibits in the Higgs-meson sector the superpotential of the type \eqref{w0}, namely
\begin{align} 
W_0 = \mu \,M_1 H_u H_d +   m\, M_1 M_2 +{\Lambda^5\over M_1 M_2}\,\,.
\label{eq D-term susybreaking}
\end{align}
From Figure \ref{fig dquark1} one can easily see that the term $M_1\, M_2$ indeed represents a closed loop (from e' to d' and back)
in the quiver diagram. 
The superpotential term $M_1 H_u H_d$ 
requires a little bit more work. Let us start at node $b$ and go to node $e'$, which describes the $H_u$ field. From node $e'$ 
we go via node $f$ to node $d'$ and pick up the meson $M_1$. To close the loop we jump to the node which is the orientifold image of $d'$, 
namely node $d$  but keeping in mind that such a jump implies reversing the orientation of the loop. Then one can close the loop
with the inclusion of $H_d$.
The  last term in \eqref{eq D-term susybreaking} is the non-perturbative ADS superpotential term \footnote{For a recent derivation of the ADS superpotential in the context of intersecting D6-branes see \cite{Akerblom:2006hx}.}. 
Note that the superpotential in 
the Higgs-meson sector is of the type \eqref{w0} discussed in Section \ref{sec D-term breaking}. There we showed that for a particular 
choice in parameter space the vacuum breaks SUSY as well as the electroweak gauge symmetry.

Analogously to the Higgs-meson sector, one can determine the superpotential containing the chiral MSSM superfields. Again that corresponds 
to finding all closed loops involving two quark and lepton fields in the quiver diagram \ref{fig dquark1}. One obtains for the quiver potential
\begin{align}
W_2 = Q_L H_u u^c+  Q_L H_d d^c+ \frac{1}{\Lambda} M_1 L H_d  e^c+ \frac{1}{\Lambda} M_1 L H_u  \nu^c \,\,.\label{w2one}
\end{align}
For simplicity we omit dimensionless couplings. This superpotential generates mass terms for quark and leptons.
Interestingly, the origin of masses for quark and leptons in (\ref{w2one}) are very different. Masses for the quarks arise from the
familiar Yukawa couplings with the MSSM Higgs fields while those for the leptons arise from the quartic couplings involving a meson field.  
More precisely, at a SUSY breaking vacuum of the type we discussed above this superpotential generates masses for the quarks of order $h_{u,d}$
 and for the leptons of order $x_{1} h_{u,d}$. In addition soft SUSY breaking masses  for the sparticles follow from non-trivial vevs of the F-fields 
 $\vec F_X$. 
  
\subsubsection*{Three generation quiver}
\begin{figure}[h]
\begin{center}
 \includegraphics[width=0.9\textwidth]{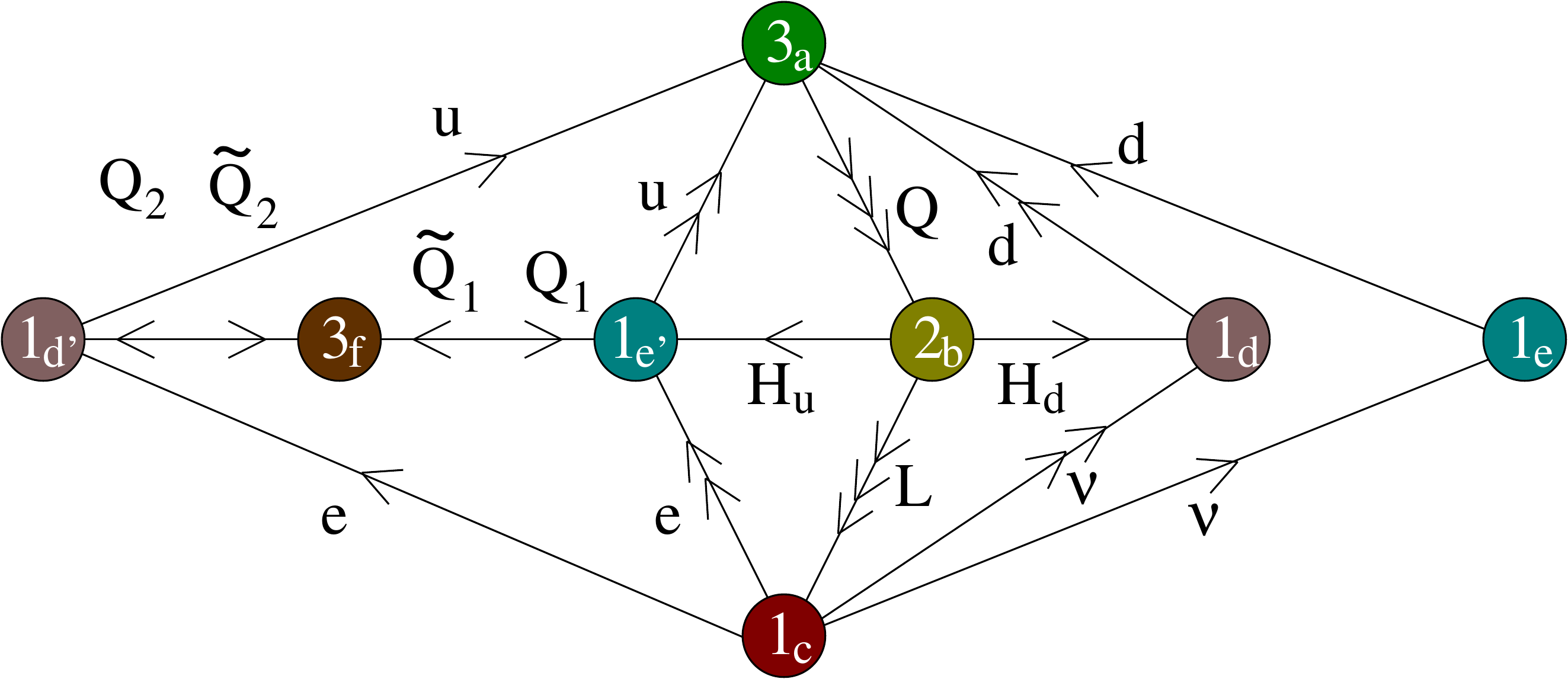}
\end{center}
\caption{\small Three generation quiver leading to SUSY breaking.}\label{fig dquark2}
\end{figure}

 The three generation quiver is displayed in Figure \ref{fig dquark2} and  given by the multiplicity choice
 \begin{align}
n=3\qquad n_u=2 \qquad n_d=2 \qquad n_e=2 \qquad n_{\nu}=2\qquad n_Y=0 \nn
%\label{eq multiplicities first example}
 \end{align} 
 which gives the MSSM spectrum
 \begin{align*} 
& \hspace{5.5mm}Q_L =3 \times (a,b) \hspace{10mm} u^c_{1,2}=2 \times (\bar a, \bar e) \hspace{11mm} u^c_{3}=1 \times (\bar a, \bar d) \\  
& \hspace{5mm} d^c_{1,2} =2 \times (\bar a,  d) \hspace{12mm} d^c_{3}=1 \times (\bar a, e) \hspace{12mm} L= 3\times (b, \bar c) \\
e^c_1 & =1 \times (c,d) \hspace{8mm}  e^c_{2,3}=2 \times (c,e) \hspace{8mm} \nu^c_1=  1\times (c, \bar e) \hspace{8mm} \nu^c_{2,3}= 2 \times (c, \bar d)\,\,. 
 \end{align*}
Let us stress that the choice of multiplicities do pass all the string consistency constraints derived in Section \ref{sec D-brane model building}.  
The Higgs-meson sector is the same as for the one generation configuration, see equation \eqref {eq Higgs meson sector D-term}. Thus it exhibits  
the same superpotential \eqref{eq D-term susybreaking} as for the one generation quiver and therefore as shown in Section \ref{sec D-term breaking} 
leads to simultaneous SUSY and electroweak gauge symmetry breaking. 

Finding all closed loops containing exactly two fields $\vec{X}$ gives the MSSM superpotential 
\begin{align}
W_2= & Q_L H_u u^c_{1,2} + Q_L \,H_d\, d^c_{1,2} + L \, H_d\, e^c_1 + L \, H_u\,  \nu^c_1
\label{pot3gen}\\& \hspace{5mm} + \frac{M_1}{\Lambda}  \Big(  Q_L H_u u^c_{3} + Q_L \,H_d\, d^c_3 + L \, H_d\, e^c_{2,3} +   L \, H_u\,  \nu^c_{2,3} \Big) \,\,. \nn
\end{align}
After the Higgs fields and $M_1$ acquire non-zero vev's this superpotential gives a mass to all the SM fields.
In (\ref{pot3gen}) the Yukawa couplings for different families have different coefficients and this might be of use to account 
for the observed mass hierarchies in the standard model. 
Note also that this quiver does not contain any R-parity violating couplings or dangerous dimension five operators.

We remark that the perturbative presence of a Dirac neutrino mass term (of the order of the SM mass scale), 
in conjunction with the presence of a large Majorana mass term 
for the right-handed neutrinos, induced by a D-instanton, might lead to a neutrino mass of the right size via the seesaw 
mechanism \cite{Ibanez:2006da,Blumenhagen:2006xt,Argurio:2007vqa,Bianchi:2007wy,Ibanez:2007tu,Ibanez:2007rs,Cvetic:2007ku,Cvetic:2007qj}.

Finally let us comment on the mass of $U(1)_d$ and $U(1)_e$. Only their sum satisfies the constraints for a massless $U(1)$. That implies that both $U(1)_d$ and $U(1)_e$ become massive via the CS-coupling. This mass crucially depends on the details of the compactification (string coupling, string mass, volume of the cycles the D-branes wrap as well as on the gauge flux living on them, etc.) \cite{Antoniadis:2002cs,Anastasopoulos:2004ga, Anastasopoulos:2006cz,Anastasopoulos:2008jt}. Here we assume that both masses are below the scale $\Lambda$, such that at the energy scale $\Lambda$ one has to treat them as an abelian gauge symmetry.

 \subsection{Second Example}
\begin{figure}[htb]
\begin{center}
 \includegraphics[width=0.9\textwidth]{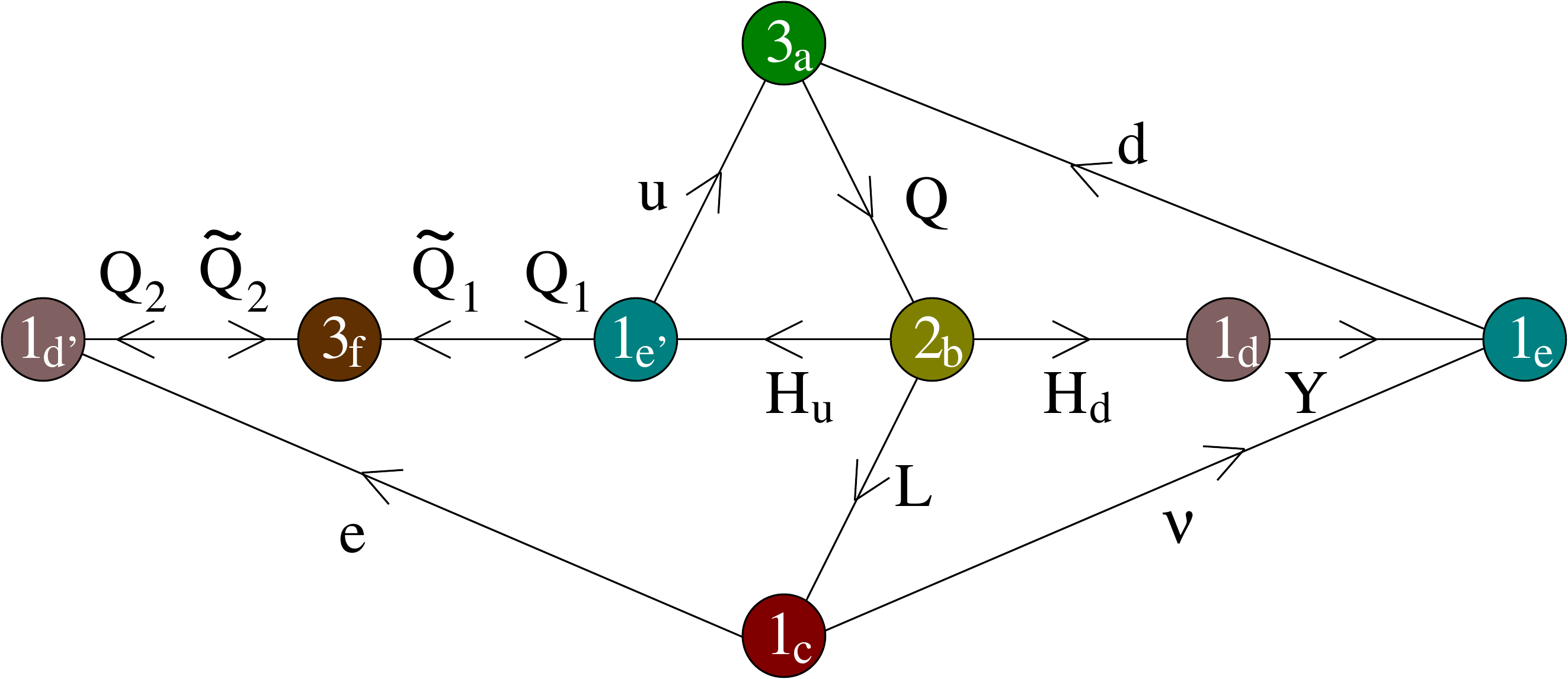}\vspace{5mm}
 \end{center}
\caption{\small One generation quiver  with additional an $Y$ leading to SUSY breaking.}\label{fig dquarkY1}
\end{figure}

 Figures \ref{fig dquarkY1} and \ref{fig dquarkY3} display the quiver diagrams for a one and three generation configuration, 
which exhibit a superpotential of the type (\ref{w02}). Let us start again with the one generation quiver.
 
 \subsubsection*{One generation quiver}
 
 For the one generation model the multiplicity numbers in Table \ref{table allmodels} are chosen
\begin{align}
n=1\qquad n_u=1 \qquad n_d=0 \qquad n_e=0 \qquad n_{\nu}=0\qquad n_Y=1 \nn
 \end{align}
and lead to the MSSM spectrum
 \begin{align} \nn
 Q_L&=(a,b) \qquad u^c=(\bar a, \bar e) \qquad d^c=(\bar a,  e) \\ \nn
  L&=(b , \bar c) \qquad \hspace{1mm} e^c =(c,d) \qquad \nu^c=(c, \bar e)
 \end{align}
 and to the Higgs-meson spectrum
  \begin{align}
 H_u&=(b,e) \hspace{5mm} H_d=(b, \bar d) \hspace{5mm} Y=(d,\bar e) \hspace{5mm}  M_1 =(d,\bar e) \hspace{5mm}M_2=(\bar d, e)
 \label{eq Higgs meson sector}
 \end{align} where again $M_1$ and $M_2$ denote the mesons after the condensation of $SU(3)_H$.   
 
 The superpotential is given by all possible loops in the quiver and takes, in the Higgs-meson sector, the form
 \begin{align}
 W_0 = \mu\, M_1 H_u H_d  + \mu_Y \, Y H_u H_d +  m\, M_1  M_2 + m_Y  \,Y M_2 + \frac{\Lambda^5}{M_1\,M_2}\,\,.
 \label{eq W_0 F-term}
 \end{align}
 This is exactly the superpotential analysed in Section \ref{sec f-term breaking}, where it has been shown that for some choice of parameters
 there exist a SUSY and electroweak gauge symmetry breaking vacuum.    
 
One moreover obtains the desired MSSM Yukawa couplings
\begin{align}
W_2= Q_L\, H_u \,u^c +  \frac{1}{\Lambda} \,Q_L \,H_d \Big( M_1 +Y \Big) \,d^c  +L \,H_d\, e^c  +L \,H_u \,\nu^c 
 \end{align}
 which gives masses to all the MSSM matter fields after $H_u$, $H_d$, $M_1$ and $Y$ acquire a non-zero vev.

\subsubsection*{Three-generation quiver}

An extension of the above discussed configuration to three families is given by the following choice for the multiplicities in Table
\ref{table allmodels}
\begin{align}
n=3\qquad n_u=2 \qquad n_d=2 \qquad n_e=2 \qquad n_{\nu}=3 \qquad n_Y=1\,\,, \nn
 \end{align}
 where again this choice of multiplicities satisfies the string consistency constraints, tadpole cancellation and masslessness of $U(1)_Y$.
With that choice one obtains the MSSM spectrum
\begin{align} \nn
Q_L &=3 \times (a,b) \hspace{10mm} u^c_{1,2}=2 \times (\bar a, \bar e) \hspace{16mm} u^c_3=1 \times (\bar a, \bar d) \\
d^c_{1,2} &=2 \times (\bar a,  d) \hspace{12.5mm} d^c_3=1 \times (\bar a, e) \hspace{16.5mm} L= 3\times (b, \bar c)\\ \nn
e^c_1 & =1 \times (c,d) \hspace{11mm}  e^c_{2,3}=2 \times (c,e) \hspace{11mm} \nu^c_{1,2,3}= 3 \times (c, \bar d)\,\,, \nn
 \end{align}
 while the Higgs-meson sector is the same as for the one generation configuration, see equation \eqref{eq Higgs meson sector}.

\begin{figure}[h]
\begin{center}
 \includegraphics[width=0.9\textwidth]{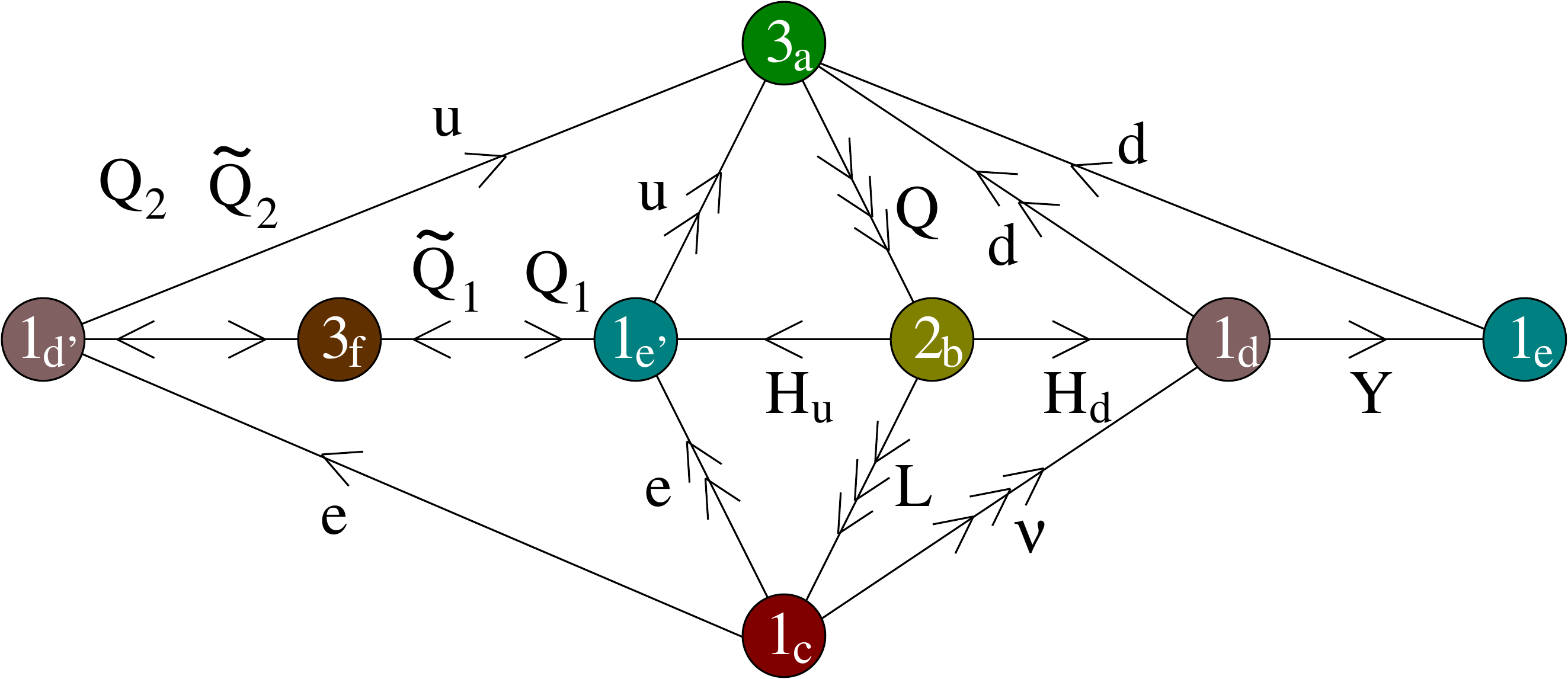}
\end{center}
\caption{\small Three generation quiver  with additional an $Y$ leading to SUSY breaking. }\label{fig dquarkY3} 
\end{figure}

The corresponding quiver is displayed in Figure \ref{fig dquarkY3} and the superpotential for the Higgs-meson sector is again given by
\eqref{eq W_0 F-term}. Thus as shown in Section \ref{sec f-term breaking} there exists a vacuum which breaks SUSY and also $SU(2) \times U(1)_Y$ gauge symmetry.
   
The MSSM superpotential is given by
\begin{align}
W_2= & Q_L H_u u^c_{1,2} + Q_L \,H_d\, d^c_{1,2} + L \, H_d\, e^c_1 
\\& \hspace{5mm} +  \frac{1}{\Lambda} \,\Big( M_1+Y\Big) \Big(  Q_L H_u u^c_3 + Q_L \,H_d\, d^c_3 + L \, H_d\, e^c_{2,3} +   L \, H_u\,  \nu^c_{1,2,3} \Big) \,\,. \nn
\end{align}  
After the Higgs-meson fields acquire a vev all SM fields get a mass. 
As in the previous example, the fact that one of the families of the right-handed quarks has a different string origin  with respect 
to the other two generations may account for the observed mass 
hierarchies in the MSSM.  As before the quiver does not contain any R-parity violating couplings or dangerous dimension five operators. Moreover, D-instantons can generate large Majorana masses for the right-handed neutrinos. Then the small neutrino masses can be explained via the seesaw mechanism.

As before we assume that the masses of the $U(1)_d$ and $U(1)_e$ induced by the CS-couplings are below the scale $\Lambda$. That forces us to treat the symmetries $U(1)_d$ and $U(1)_e$ as gauge symmetries at the energy scale $\Lambda$.

\section{Summary}

We discussed two extensions of the MSSM which lead, for a large region in parameter space, to SUSY and electroweak gauge symmetry breaking. 
Both these extensions contain a hidden $SU(3)_H$ that condensates via the generation of an ADS superpotential. The condensates (mesons) couple to the  
Higgs sector which mediates the SUSY  and electroweak gauge symmetry breaking to MSSM matter content. 
In Section \ref{sec Susybreaking} we show explicitly that, depending on the region in parameter space, these extensions can give rise to 
different vacua, that 
do or do not break SUSY and/or  electroweak gauge symmetry.  There are large region in parameter space for which both symmetries are broken.

Later  in  Section \ref{sec consistent string realizations} we present local D-brane configurations which satisfy severe string consistency constraints 
and mimic the previously discussed field theory setups. They exhibit the required superpotential to break SUSY and  electroweak gauge symmetry. 
Moreover, all MSSM Yukawa couplings are realized and these configurations can naturally explain some of the observed mass hierarchies of the MSSM. 

In this work we assumed that the closed string sector is stabilized thus we ignore all closed string dynamics. Specifically all Yukawa, gauge couplings 
and Fayet-Iliopoulos terms are input parameters. It would be nice to find a global realization of these local D-brane configurations in which one can 
study whether moduli stabilization indeed give the Yukawa and gauge couplings, as well as Fayet-Iliopoulos terms in a range that eventually leads to 
SUSY and gauge symmetry breaking.

 \vskip 2cm \noindent {\large {\bf Acknowledgments}} \vskip 0.2cm
The authors wish to thank P. Anastasopoulos, M. Bianchi, E. Dudas and M. Serone for useful discussions.
This work was partially supported by
the ERC Advanced Grant n.226455 {\it ``Superfields''}, by the
Italian MIUR-PRIN contract 20075ATT78 and by the NATO grant
PST.CLG.978785.

\noindent \vskip 1cm

%\newpage
%\bibliographystyle{abe}
%\bibliography{refsusybreak}
\providecommand{\href}[2]{#2}\begingroup\raggedright\endgroup

\end{document}